\documentclass[conference]{IEEEtran}

\usepackage{amsmath,amssymb}
\usepackage{subfigure}
\usepackage{graphicx,graphics,color,psfrag}
\usepackage{cite,balance}
\usepackage{caption}
\allowdisplaybreaks
\usepackage{accents}
\usepackage{amsthm}
\usepackage{bm}
\usepackage[english]{babel}
\usepackage{multirow}
\usepackage{enumerate}
\usepackage{cases}
\usepackage{stfloats}
\usepackage{dsfont}
\usepackage{color,soul}
\usepackage{amsfonts}
\usepackage{cite,graphicx,amsmath,amssymb}
\usepackage{subfigure}
\usepackage{fancyhdr}
\usepackage{hhline}
\usepackage{graphicx}
\usepackage{array,color}
\usepackage{booktabs}
\usepackage{diagbox}
\usepackage{indentfirst}
\usepackage{url}
\usepackage[ruled,vlined]{algorithm2e}
\usepackage{makecell}
\usepackage{soul}

\usepackage[a4paper, top=1.8cm, bottom=4.21cm, left=1.4cm, right=1.4cm]{geometry}

\usepackage{amsmath}

\usepackage{lipsum}

\theoremstyle{definition}

\usepackage{lipsum}

\begin{document}
	\title{Fast and Accurate Cooperative Radio Map Estimation Enabled by GAN}
	
	\IEEEoverridecommandlockouts{
		\author{\IEEEauthorblockN {Zezhong Zhang\IEEEauthorrefmark{1}\IEEEauthorrefmark{2},
				Guangxu Zhu\IEEEauthorrefmark{3}, Junting Chen\IEEEauthorrefmark{1}\IEEEauthorrefmark{2}, and Shuguang Cui\IEEEauthorrefmark{1}\IEEEauthorrefmark{2}\IEEEauthorrefmark{4}}
			
			\IEEEauthorblockA{
				\IEEEauthorrefmark{1}School of Science and Engineering (SSE) and \IEEEauthorrefmark{2}Future Network of Intelligent Institute (FNii), \\The Chinese University of Hong Kong (Shenzhen), Shenzhen, China\\
				\IEEEauthorrefmark{3}Shenzhen Research Institute of Big Data, Shenzhen, China\\
				\IEEEauthorrefmark{4}Pengcheng Lab, Shenzhen, China}	
	}}
	\date{}
	\maketitle

\begin{abstract}
In the 6G era, real-time radio resource monitoring and management are urged to support diverse wireless-empowered applications. This calls for fast and accurate estimation on the distribution of the radio resources, which is usually represented by the spatial signal power strength over the geographical environment, known as a radio map. In this paper, we present a cooperative radio map estimation (CRME) approach enabled by the generative adversarial network (GAN), called as GAN-CRME, which features fast and accurate radio map estimation without the transmitters' information. The radio map is inferred by exploiting the interaction between distributed received signal strength (RSS) measurements at mobile users and the geographical map using a deep neural network estimator, resulting in low data-acquisition cost and computational complexity. Moreover, a GAN-based learning algorithm is proposed to boost the inference capability of the deep neural network estimator by exploiting the power of generative AI. Simulation results showcase that the proposed GAN-CRME is even capable of coarse error-correction when the geographical map information is inaccurate.

\end{abstract}

\section{Introduction}
	
Riding the wave of mobile edge computing and \emph{artificial intelligence} (AI), recent years have witnessed an exponential growth of distributed data generated by diverse applications at the network edge \cite{CS1, Dingzhu2}. This calls for ubiquitous storage, communication and computation to fulfill the task-specific requirements \cite{Dingzhu1}, which vary according to the users' demands and surrounding environments. For example, stable signal coverage along the trajectory is essential for near-ground UAV services which require reliable connection. To provide on-demand support for diverse applications, the key is to achieve real-time awareness of the environment. 
Particularly, the distribution information of the radio resources is a key characteristic of the environment and needs be estimated accurately in real-time for efficient radio resource management \cite{Application1, VFL}. The distribution of the radio resources is usually represented by the spatial signal power strength over the geographical environment, known as a radio map after visualization \cite{Definition1,Definition2}. Therefore, accurate and fast \emph{radio map estimation} (RME) designs are crucial to the environment-aware intelligent radio resource management in 6G networks \cite{Application1, Definition1}. 

Since the inception of the radio resource allocation concept, the RME methods have continued to evolve in tandem with advancements in signal processing and machine learning techniques. Conventional approaches using data-driven interpolation \cite{kriging2,kernel,tensorcompletion,Junting1} or model-based ray-tracing methods \cite{ray} are unable to achieve high accuracy and speed simultaneously \cite{kriging2, Junting2, ray}. Recently, deep learning based methods that approximates the ray-tracing models with its universal function approximation capability attract much attention due to their potential to exploit the geographical environment information \cite{autoencoder,Unet}. However, the existing approaches are not fully ready for the implementation in practical systems due to the following two reasons. Firstly, all existing RME methods with high accuracy, e.g., ray-tracing methods and the RadioUNet \cite{Unet}, rely on the physical model of signal propagation and thereby require the transmitters' information, i.e., their location and transmit power. Since the transmitters can be any base stations or mobile users, it is impossible to guarantee real-time acquisition on the transmitters' information, making these methods infeasible. The second challenge is the generalization capability. The mismatch between the practical and simulated scenarios, i.e., the unexpected perturbation or imperfection on the geographical environment information, may distort the estimation results. 

To handle the above challenges, an alternative solution is provided in this work to replace the transmitters' information with distributed \emph{received signal strength} (RSS) samples at mobiles users, called as \emph{cooperative RME} (CRME). Distributed RSS samples are easy to obtain and most importantly endow with the capability of error-correction in the presence of inaccurate geographical map information, since they are sampled from the ground-truth radio map. Moreover, to harness the recent advancement of generative AI, a conditional \emph{generative adversarial network} (GAN) based learning algorithm is further proposed to improve the accuracy where only sparse RSS samples over the region of interest is available, giving the name of the proposed approach as GAN-CRME. With a limited number of RSS samples, GAN-CRME is able to achieve favorable accuracy compared with the state-of-the-art methods which all require the transmitters’ information. Our design and contributions are summarized as follows.

In this work, we develop a fast and environment-aware RME approach so that it can be applied in different environments with the assist of corresponding geographical map. 
The GAN-CRME approach has three key features as elaborated below.

\begin{itemize}
\item {\bf Real-time RME: }  
Firstly, a \emph{deep neural network} (DNN) estimator is used, whose low computational complexity guarantees fast inference. Moreover, the CRME framework also features low data-acquisition cost, where the high-resolution geographical map can be downloaded from the Internet in prior and the RSS measurements at mobile users can be directly uploaded in the uplink phase. 	
		
\item {\bf GAN-enabled Accuracy Improvement:} 
The CRME framework faces the key challenge of low accuracy since the embedded physical model is weak. To improve the quality of the estimated radio map, we propose a GAN-based learning algorithm to harness the power of generative AI. The DNN estimator is trained together with a discriminator using correlated loss functions. This establishes an adversarial training process, where the inference capability of the DNN estimator is greatly enhanced. After training, the DNN estimator is able to fully exploit the interaction between the distributed RSS samples and the geographical map, resulting in high accuracy and even the error-correction capability. \emph{To our best knowledge, this is the first work which provides a solution to real-time RME by incorporating distributed RSS samples and the geographical map information.}

\item {\bf High Model Generalization Capability:} The model generalization capability in GAN-CRME is effectively improved by the following two designs. Firstly, a random number of buildings are removed from the input geographical map to mimic the case with inaccurate environment information. Secondly, the GAN-based learning algorithm helps exploit the deep features embedded in the RSS measurements, which are sampled from the ground-truth radio map, to correct errors brought in by the inaccurate geographical map and restrain the performance degradation in real scenarios. 

	\end{itemize}
	

\section{System Model}\label{SystemModel}
In this section, we present the model of a wireless communication system and formulate the CRME problems. 
	

\subsection{Communication Model}\label{broadband}
We consider an $R_1 \times R_2 \ (R_1,\ R_2 \in \mathbb{N}^+)$ rectangular area with a number of mobile users, where $K_t$ among them are able to measure the RSS and upload it together with their location information to the server at the current time slot $t$. The transmitters can be the base stations or mobile users with unknown numbers and locations. The transmitter set is denoted as $\mathcal{T}_t$. According to the signal propagation model, the received signal $Y_j$ at device $j$ can be written as
\begin{align}\label{PropagateModel}
    Y_j = \sum_{i \in \mathcal{T}_t}\sqrt{P_i \cdot \rho_{i,j}}h_{i,j}X_i + Z_j,
\end{align}
where $\rho_{i,j}$ and $h_{i,j}\!\sim\!\mathcal{CN}(0,1)$ denote the large-scale and small-scale fading coefficients of the channel between transmitter $i$ and device $j$, $X_i\!\sim\!\mathcal{CN}(0,1)$ is the normalized signal emitted from transmitter $i$, $P_i$ is the transmit power, and $Z_j\!\sim\!\mathcal{CN}(0,\sigma_z^2)$ is the Gaussian noise at the receiver. The time slot subscript $t$ is omitted for simplicity in the rest part. Then according to Eq. \eqref{PropagateModel}, the average RSS is calculated as 
\begin{align}\label{rssUserj}
    \mathbb{E}[\|Y_j\|^2] = \sum_{i \in \mathcal{T}} P_i \cdot \rho_{i,j} + \sigma_j^2.
\end{align}
The large-scale fading coefficient $\rho_{i,j} = \rho_{i,j}^{sh}\cdot \text{PLoss}_{i,j}$, which is the product of the shadowing effect $\rho_{i,j}^{sh}$ and the pathloss $\text{PLoss}_{i,j}$, is determined by the geographical environment. We denote the location of device $j$ as $(x_j, y_j)$, where $x_j$ and $y_j$ are integers within ranges $[0, R_1]$ and $[0, R_2]$.  According to the above elaboration, the average RSS in \eqref{rssUserj} can be written as 
\begin{align}\label{functionUserj}
    \mathbb{E}[\|Y_j\|^2] = f(\mathbf{T},\boldsymbol{\Phi}^\text{map})_{(x_j,y_j)}+\sigma_j^2,
\end{align}
where $\mathbf{T}\in\mathbb{R}^{R_1\times R_2}$ is a matrix representing the transmitters' information, given as $\mathbf{T}_{(x_i,y_i)} = P_i$ if $i\in\mathcal{T}$ and $\mathbf{T}_{(x_i,y_i)} = 0$ otherwise. The above equation indicates that the average RSS at an arbitrary location can be expressed as a function of the transmitters' information $\mathbf{T}$ and the ground-truth geographical map $\boldsymbol{\Phi}^\text{map} \in \mathbb{R}^{R_1\times R_2}$ plus a noise power term. The average noise power $\sigma_j^2$ at the receiver is a time-invariant constant and therefore the noise power map can be represented by a constant matrix $\mathbf{\Sigma}\in\mathbb{R}^{R_1\times R_2}$.  Then according to Eq. \eqref{functionUserj}, the average RSS across the whole region, i.e., the radio map $\mathbf{P} \in \mathbb{R}^{R_1 \times R_2}$, can be described as
\begin{align}\label{radioMap}
    \mathbf{P} = f(\mathbf{T}, \mathbf{\Phi}^\text{map})+\mathbf{\Sigma},
\end{align}
where the average RSS at the location of device $j$ is represented as
\begin{align}\label{PointwiseRadioMap}
	\mathbf{P}_{(x_j,y_j)} = f(\mathbf{T}, \mathbf{\Phi}^\text{map})_{(x_j,y_j)}+\sigma_j^2.
\end{align}

\subsection{CRME Model}\label{CRMEModel}

As elaborated above, the radio map $\mathbf{P}$ is modeled as a function of the transmitter information $\mathbf{T}$ and the geographical environment $\boldsymbol{\Phi^\text{map}}$. However, the transmitters' information is usually unavailable in wireless networks and thereby unsuitable for the real-time RME. In the absence of the transmitters' information, we propose the idea to exploit the interaction between the distributed RSS samples at mobile users and the geographical map for RME, which relies on the cooperation of the distributed mobile users and is called as \emph{cooperative RME} (CRME). The feasibility of the idea proposed above is based on the fact that the transmitters' information are embedded in the RSS information as deep features, giving the opportunity to infer the radio map by exploiting the interactions between the RSS samples and the geographical map.  
In the following, a mathematical description on the embedded physical model of the proposed CRME is provided, where two typical cases are considered with accurate and inaccurate geographical maps.

\subsubsection{Case 1: CRME with Accurate Geographical Map} 
The relation between the transmitters and the resultant radio map is depicted by Eq. \eqref{radioMap}. Inversely, in view of the sparsity of the transmitters, their information is also uniquely determined by distributed RSS samples given the geographical map. Suppose that there are $K$ active users available at the current time slot, and the active user set is denoted as $\mathcal{K}$. An estimate of the transmitters' information can be represented by an undetermined function $H(\cdot)$ as
\begin{align}\label{replaceTx}
	\tilde{\mathbf{T}} = H\left( \mathbf{R}, \mathbf{\Phi}^\text{map}\right).
\end{align}
The matrix $\mathbf{R}\in\mathbb{R}^{R_1\times R_2}$ represents the RSS samples over the network, where $\mathbf{R}_{(x_j,y_j)} = \mathbf{P}_{(x_j,y_j)}$ if $j\in\mathcal{K}$ and $\mathbf{R}_{(x_j,y_j)} = 0$ otherwise. Based on Eq. \eqref{radioMap} and \eqref{replaceTx}, an estimate of the radio map is modeled as 
\begin{align}\label{RMExpression}
   \tilde{\mathbf{P}} = f\left( H\left(\mathbf{R}, \mathbf{\Phi}^\text{map}\right),  \mathbf{\Phi}^\text{map}\right) +\mathbf{\Sigma} = F_1\left( \mathbf{R}, \mathbf{\Phi}^\text{map} \right).
\end{align}

\subsubsection{Case 2: CRME with Inaccurate Geographical Map}

We also consider a more realistic scenario where the geographical map information is inaccurate, e.g., there are some buildings or temporarily parked cars in the region of our concern but missing in the geographical map that we can obtain. 
Following the CRME model \eqref{RMExpression} in Case 1, the error in the geographical map will distort the resultant radio map estimate  inevitably. However, note that the input RSS samples $\mathbf{R}$ are a product of interactions between the signal emitted from the transmitters and the ground-truth geographical map, given as 
\begin{align}
	\mathbf{R}_{(x_j,y_j)} = f(\mathbf{T}, \mathbf{\Phi}^\text{map})_{(x_j,y_j)}+\sigma_j^2,
\end{align}
according to Eq. \eqref{PointwiseRadioMap}. Then the ground-truth geographical map can be considered as deep features embedded in the RSS samples. Therefore, once we have enough RSS samples, it is possible to exploit the deep features to correct partial errors brought in by the inaccurate geographical map $\tilde{\mathbf{\Phi}}^\text{map}$. According to the elaboration above, in this case the given information is still sufficient for CRME and the solution follows the form
\begin{align}
	\tilde{\mathbf{P}} = F_2\left( \mathbf{R}, \tilde{\mathbf{\Phi}}^\text{map} \right), 
\end{align}
It is also worthwhile to notice that $F_2(\cdot)$ is different from $F_1(\cdot)$, where the importance of the RSS information in $F_2(\cdot)$ is higher for error-correction.

\subsection{Problem Formulation}\label{problemFormulation}
\subsubsection{Case 1} 
Since the signal propagation process is complicated, and the function $H(\cdot)$ is undetermined, the expression of the function $F_1(\cdot)$ in Eq. \eqref{RMExpression} is difficult to obtain. However, in view of the fact that DNNs are capable of approximating arbitrary functions, we propose to substitute function $F_1(\cdot)$ with a DNN generator $G_1$ with parameters $\boldsymbol{\Theta}_{G_1}$. During the training process, in each round we try to optimize the generator to minimize the distance between the generated image ${G_1}(\mathbf{X};\boldsymbol{\Theta}_{G_1})$ and the ground-truth label $\mathbf{P}$, given as
\begin{align}
    (\mathbf{P1}) \qquad \min\limits_{\boldsymbol{\Theta}_{G_1}} \sum\limits_{(\mathbf{X},\mathbf{P})\in\mathcal{D}_1} \text{Dist}\left( {G_1}(\mathbf{X}; \boldsymbol{\Theta}_{G_1}), \mathbf{P} \right),
\end{align}
where $\mathcal{D}_1$ is the training dataset of $N$ data samples, $\mathbf{X} = \left\{\mathbf{R}, \mathbf{\Phi}^\text{map} \right\}$ is the feature vector in a data sample, and the distance function $\text{Dist}(\cdot)$ is defined later in Sec. \ref{learningAlgo} according to the proposed GAN-based algorithm.

\subsubsection{Case 2} 
 Similar to the Case 1, the optimization problem to solve in each round is formulated as
\begin{align}
    (\mathbf{P2}) \qquad \min\limits_{\boldsymbol{\Theta}_{G_2}} \sum\limits_{(\tilde{\mathbf{X}},\mathbf{P})\in\mathcal{D}_2} \text{Dist}\left( {G_2}(\tilde{\mathbf{X}}; \boldsymbol{\Theta}_{G_2}), \mathbf{P} \right),    
\end{align}
where $\tilde{\mathbf{X}} = \left\{\mathbf{R}, \tilde{\mathbf{\Phi}}^\text{map} \right\}$ is the feature vector with error.

With similar forms, the above two CRME problems are solved using the same approach in this work. Therefore, we take Case 1 as an example for illustration by default and the subscript of the generator is omitted for convenience.


\section{GAN-enabled Cooperative Radio Map Estimation}\label{GANAlgorithm}
In this section, we first present the pre-processing operation on dataset, and then describe the learning architecture, model setting and the proposed GAN-based learning algorithm.


\subsection{Pre-processing}\label{preprocess} 
We propose to use a UNet model as the DNN estimator for an image-to-image estimation in this work. Therefore, the data samples need to be transformed into the image form in the first stage. For each data sample $(\mathbf{X}_\ell, \mathbf{P}_\ell)$ in dataset $\mathcal{D}$ corresponding to environment $\ell$, the power levels in matrices $\mathbf{R}_\ell$ and $\mathbf{P}_\ell$ are firstly transformed into the grey levels using logarithm operation, normalization, and quantization \cite{Unet}. Here we use the same representations of $\mathbf{R}_\ell$ and $\mathbf{P}_\ell$ for the original and image-form data for simplicity. The feature vector is made from the original one as a 2-channel image $\mathbf{X}^\text{img}_\ell=[\mathbf{R}_\ell, \mathbf{\Phi}^\text{map}_\ell]$ by concatenation. In this step, we construct a new dataset $\mathcal{D}^\text{img}$ where each data sample is presented as $(\mathbf{X}^\text{img}_\ell, \mathbf{P}_\ell)$.

\subsection{Conditional GAN Architecture}
\begin{figure*}[t]\vspace{-7mm}
			\centering
			\subfigure[Loss calculation based on the real data samples.]{\includegraphics[height=3cm]{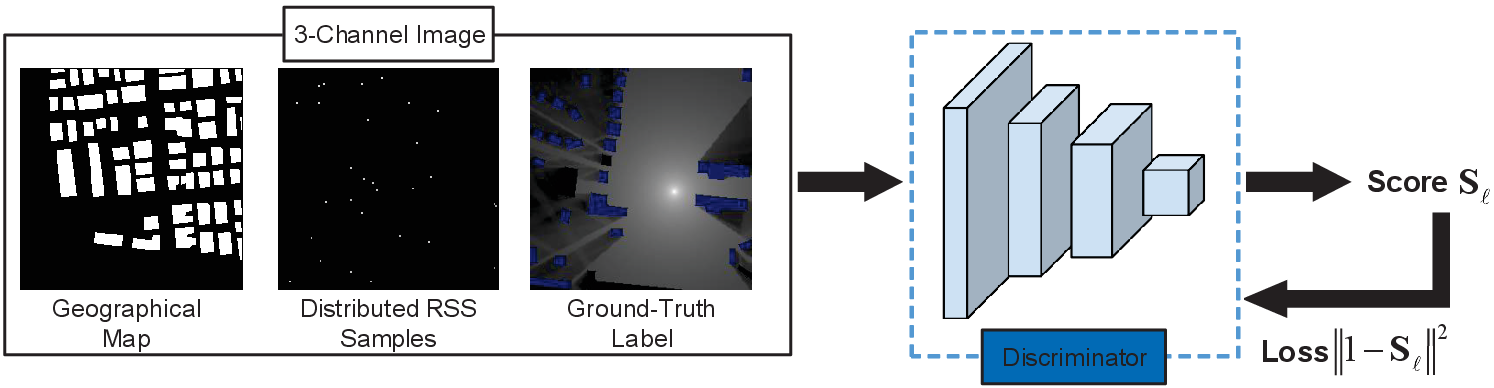}}
			\subfigure[Fake image generation and the corresponding loss calculation.]{\includegraphics[height=4cm]{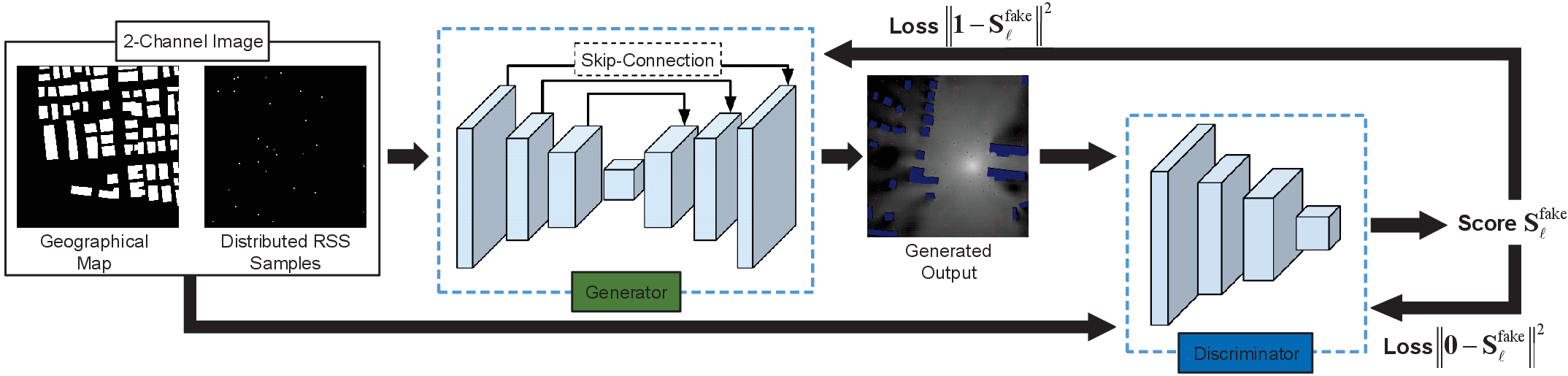}}
		\caption{Diagram of the proposed GAN-CRME training process.}\label{CGAN}\vspace{-4mm}
\end{figure*}

Conditional GAN consists of a generator and a discriminator inherited from the standard GAN architecture. The generator is the desired DNN model for generating a radio map estimate based on the input materials. The discriminator is a network who takes charge of judging how ``real" the generated image is by measuring the quality of the input image based on its deep features.  While different from the standard GAN architecture where only the generated image and the label are fed to the discriminator, in conditional GAN the discriminator distinguishes between two joint distributions: 1) the joint distribution of the source image and the label $P(\mathbf{X}, \mathbf{P})$, and 2) the joint distribution of the source image and the generated image $P(\mathbf{X}, \tilde{\mathbf{P}})$. The generated image is also considered as a fake image in GAN, given as $\mathbf{P}^\text{fake}_\ell = \tilde{\mathbf{P}}$. The functionalities of the generator and discriminator are described below. The flow diagram of conditional GAN is also shown in Fig. \ref{CGAN}.
\begin{itemize}
    \item \textbf{Generator:} For each data sample $(\mathbf{X}^\text{img}_\ell, \mathbf{P}_\ell)$ in the dataset, the generator takes the feature vector $\mathbf{X}^\text{img}_\ell$ as input and outputs a radio map estimate $\mathbf{P}^\text{fake}_\ell$. Variants of \emph{convolutional neural network} (CNNs) are usually adopted as the generator for image-relevant tasks.

    \item \textbf{Discriminator:} In conditional GAN, the discriminator calculates a score for each real or generated data sample, i.e. $(\mathbf{X}^\text{img}_\ell, \mathbf{P}_\ell)$ or $(\mathbf{X}^\text{img}_\ell, \mathbf{P}^\text{fake}_\ell)$, to measure the quality and rationality of the data sample. The discriminator is also usually a neural network.
\end{itemize}
Note that although the generator and discriminator are jointly trained in the training phase, only the generator is required in the inference phase. 

\subsection{Models for the Generator and Discriminator}
As shown in Fig. \ref{CGAN}, we use a UNet model for the generator, which follows the autoencoder structure with additional skip-connection links. The input materials are firstly fed into a decoder for deep feature extraction, followed by an encoder for producing the desired output. The skip-connection delivers original features to the encoder, which helps to fuse features from different levels and result in higher quality of the output image. For the discriminator, a CNN model is used to measure the image quality based on its deep features. 

\begin{algorithm}[t!]
	\caption{Model Training under the Conditional GAN Architecture.\label{alg:algorithm2}}	
	\KwIn{Dataset $\mathcal{D}^\text{img}$, generator parameter $\boldsymbol{\Theta}_G$ and discriminator parameter $\boldsymbol{\Theta}_D$.}
	\KwOut{Generator parameter $\boldsymbol{\Theta}_G$ and discriminator parameter $\boldsymbol{\Theta}_D$.}
	
	{\bf Initialize} $\boldsymbol{\Theta}_G$, $\boldsymbol{\Theta}_D$, $\text{Epoch} = 0$;
	
	\While{$\text{Epoch} < N_\text{stop}$}{
		Let $\mathcal{L}_G = 0$, $\mathcal{L}_D = 0$;
		
		\For{$(\mathbf{X}^\text{img}_\ell, \mathbf{P}_\ell) \in \mathcal{D}^\text{img}$}{
			
			Calculate $\mathbf{P}^\text{fake}_\ell = G(\mathbf{X}^\text{img}_\ell)$;
			
			Feed $(\mathbf{X}^\text{img}_\ell, \mathbf{P}_\ell)$ into the discriminator;
			
			Calculate a score $\mathbf{S}_\ell = D\left( (\mathbf{X}^\text{img}_\ell, \mathbf{P}_\ell) \right)$;
			
			Feed $(\mathbf{X}^\text{img}_\ell, \mathbf{P}^\text{fake}_\ell)$ into the discriminator;
			
			Calculate a score $\mathbf{S}^\text{fake}_\ell = D\left( (\mathbf{X}^\text{img}_\ell, \mathbf{P}^\text{fake}_\ell) \right)$;
			
			\tcp{Calculate the loss for the discriminator}
			
			$\mathcal{L}^{(\ell)}_D = \|\mathbf{1}-\mathbf{S}_\ell\|^2 + \|\mathbf{0}-\mathbf{S}^\text{fake}_\ell\|^2$;
			
			Let $\mathcal{L}_D \leftarrow \mathcal{L}_D + \mathcal{L}^{(\ell)}_D$;
			
			\tcp{Calculate the loss for the generator}
			
			$\mathcal{L}^{(\ell)}_G = \|\mathbf{1}-\mathbf{S}^\text{fake}_\ell\|^2 + \lambda\| \mathbf{P}_\ell - \mathbf{P}^\text{fake}_\ell\|^2$;
			
			Let $\mathcal{L}_G \leftarrow \mathcal{L}_G + \mathcal{L}^{(\ell)}_G$;
			
		}
		
		\tcp{Train the Models}
		
		$\boldsymbol{\Theta}_D \leftarrow \boldsymbol{\Theta}_D - \eta_D \nabla \mathcal{L}_D$;
		
		$\boldsymbol{\Theta}_G \leftarrow \boldsymbol{\Theta}_G - \eta_G \nabla \mathcal{L}_G$;
		
		Let $\text{Epoch} \leftarrow \text{Epoch} + 1$;
		
	}\label{Training}
\end{algorithm}

\subsection{GAN-based Learning Algorithm}\label{learningAlgo}
To enable the generator and discriminator to produce high-quality radio maps and measure their rationality correctly, the training algorithm is given as Algorithm \ref{Training}. As can be seen, the discriminator gives both ``real" label $\mathbf{P}_\ell$ and ``fake" generated output $\mathbf{P}_\ell^\text{fake}$ scores, i.e., $\mathbf{S}_\ell$ and $\mathbf{S}_\ell^\text{fake}$, respectively. Then in each training round, we first train the discriminator with a frozen generator. By minimizing the loss function 
\begin{align}\label{loss_D}
	\mathcal{L}^{(\ell)}_D = \|\mathbf{1}-\mathbf{S}_\ell\|^2 + \|\mathbf{0}-\mathbf{S}^\text{fake}_\ell\|^2,
\end{align}
the discriminator is trained to enlarge the distance between $\mathbf{P}_\ell^\text{fake}$ and $\mathbf{P}_\ell$. After training the discriminator, the generator is trained with a frozen discriminator to minimize the distance function $\text{Dist}(\cdot)$ in Sec. \ref{problemFormulation}. The distance function is designed as the loss function
\begin{align}\label{loss_G}
\mathcal{L}^{(\ell)}_G = \|\mathbf{1}-\mathbf{S}^\text{fake}_\ell\|^2 + \lambda\| \mathbf{P}_\ell - \mathbf{P}^\text{fake}_\ell\|^2,
\end{align}
where $\lambda$ is a weight coefficient. By training the generator, the score difference and pixel-wise distance between $\mathbf{P}_\ell^\text{fake}$ and $\mathbf{P}_\ell$ are reduced. Therefore, the loss functions in \eqref{loss_D} and \eqref{loss_G} are designed to create an adversarial relation between the training processes of the generator and discriminator. The capabilities of both networks are enhanced through the adversarial training process as described in the following. As the generator is trained to output images with higher quality, the loss in \eqref{loss_D} becomes larger, resulting in a more intelligent discriminator. The parameters in the loss function of the generator \eqref{loss_G} is updated in tandem with the discriminator, which in return forces the generator to provide more ``realistic" estimation results. In the next section, we show that high-accuracy CRME can be achieved with the GAN-based learning algorithm.


\section{Experiment Results}
In this section, detailed designs of the models and training settings are provided for both cases with accurate or inaccurate geographical map information. Simulation results are given to verify the effectiveness of our proposed algorithm. We also provide a comparison between GAN-CRME and the state-of-the-art RadioUNet approach \cite{Unet}, which is an insightful work to develop a DNN-based solution with favorable accuracy. 
The RadioUNet approach relies on the transmitter's information, which is thereby a low-complexity replacement of the model-based ray-tracing methods. In this section, we show that GAN-CRME is able to achieve a higher accuracy even without the transmitter's information. Another reason for the benchmark choice is that both methods are trained and tested on the same open-source dataset. This ensures the consistency and fairness.

\subsection{Dataset}
To show the effectiveness of the proposed GAN-CRME and facilitate the comparison with the state-of-the-art methods, we use the \emph{RadioMapSeer} Dataset  at \url{http://RadioMapSeer.github.io} built by researchers who designed the RadioUNet \cite{Unet}. Each data sample is composed of features containing the transmitter's location and transmit power, and a geographical map. The labels are produced using \emph{WinProp}, a ray-tracing software at $5.9$ GHz with a $10$ MHz bandwidth. To adjust to the settings of the two cases in Sec. \ref{problemFormulation}, two types of datasets are further constructed as elaborated below.

\begin{itemize}
    \item \textbf{Standard Dataset:} For each data sample, the feature vector is a 2-channel image by concatenating the ground-truth geographical map and a random number of RSS values sampled from the label, corresponding to the description in \ref{preprocess}.

    \item \textbf{Flawed Datasets:} The Flawed Datasets are directly replicated from the Standard Dataset, while a certain/random number of buildings are removed from the input geographical map to produce some errors in the feature vector for each data sample, which thereby leads to a mismatch between the input feature vector and the ground-truth label. This is to mimic the case in practice since the perfect environment information is always unavailable due to dynamic and random perturbations. Then after training, the model is expected to be able to infer the desired radio map with inaccurate geographical map information.

\end{itemize}

\begin{figure}[t]
			\centering
			\subfigure[Structure of the UNet model as the generator.]{\includegraphics[height=3.7cm]{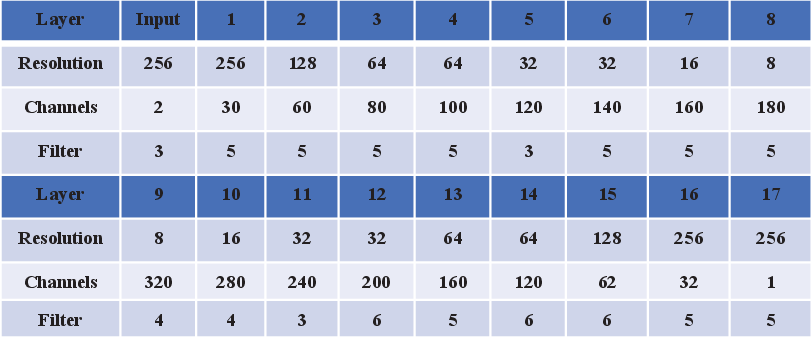}}
			\subfigure[Structure of the CNN model as the discriminator.]{\includegraphics[height=2.1cm]{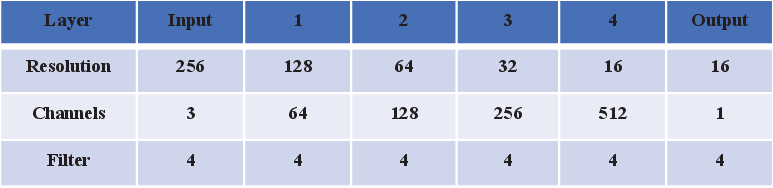}}
		\caption{Models used in the GAN-CRME architecture. Resolution is the number of pixels of the
image in each feature channel along the $x$, $y$ axis. Filter is the number of pixels of each filter kernel along the $x$, $y$ axis. The input layer is concatenated in the last two layers.}\label{ModelSetting}\vspace{-5mm}
\end{figure}

\subsection{Performance Evaluation}
In simulations, a 17-layer UNet and a 5-layer CNN are used as the generator and discriminator, with detailed structures presented in Fig. \ref{ModelSetting}. The models are shared for the two following cases. 
\subsubsection{CRME for Case 1}
\begin{figure}[t]
	\centering
	\includegraphics[width=8.8cm]{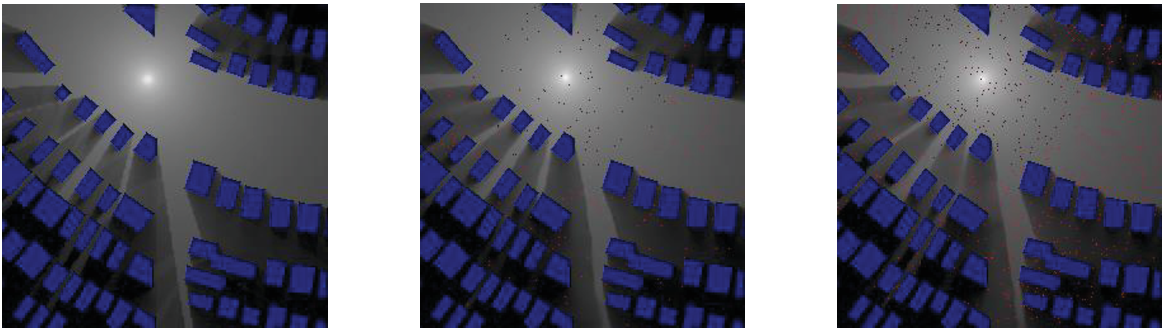}
	\caption{A radio map label (left) and estimation results using $300$ (middle) and $1000$ (right) RSS samples over the Standard Dataset.}\label{DPM_Complete_Comparison}\vspace{-3mm}
\end{figure}

To mimic the scenario with accurate environment information, we use the Standard Dataset to verify the model's capability of generating high-quality radio maps. To show the effectiveness of the proposed GAN-CRME, here we assume that the number of users $K$ follows a uniform distribution within a $256\times256$ $m^2$ area, where $R_1 = R_2 = 256$ with a 1 meter pixel length. Two uniform distributions, i.e., $U[250, 350]$ and $U[950, 1050]$ with expectation $300$ and $1000$, are used to show estimation accuracies with different number of RSS samples. The estimation results are shown in Fig. \ref{DPM_Complete_Comparison}.

\subsubsection{CRME for Case 2}\label{SimImperfectGeo}
\begin{figure}[t]
	\centering
	\includegraphics[width=8.8cm]{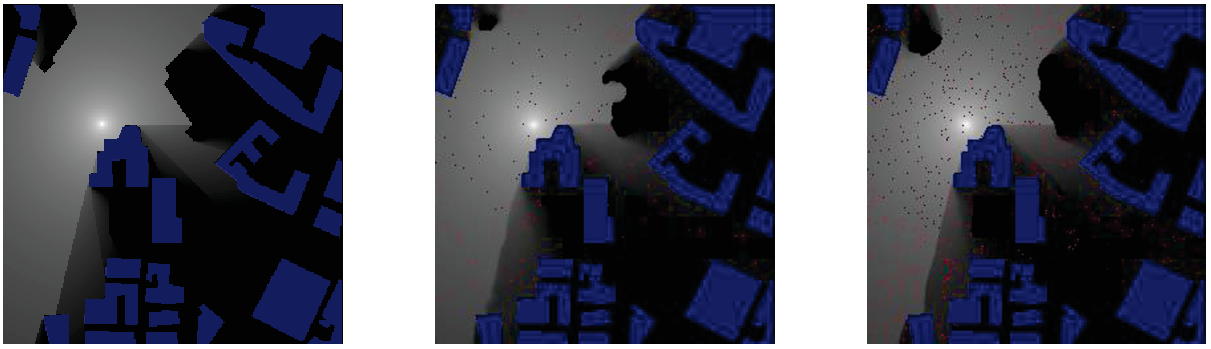}
	\caption{Illustration of the error-correction phenomenon by using $300$ RSS samples (middle) and $1000$ RSS samples (right) in GAN-CRME. The ground-truth label is presented in the left hand side.}\label{ErrorCorrection}
\end{figure}

To mimic the scenario with inaccurate environment information, we train the model using the Flawed Datasets. The well-trained model gains the error-correction capability, where some missing buildings are recovered in the final estimation results. This is attributed to the following two designs. Firstly, distributed RSS data sampled from the ground-truth radio map are used as input, offering the opportunity for the DNN to extract deep features of the ground-truth geographical environment. Secondly, to detect contradiction between the flawed geographical features and deep features extracted from the RSS samples, and further recover the missing buildings, it requires strong capability of logistic reasoning, which can be achieved by exploiting the generative AI technique, i.e., the GAN-based learning algorithm. Moreover, the error-correction capability is weak when the number of users is small since the DNN cannot collect enough deep features from the RSS samples. Therefore, the error-correction phenomenon becomes clear when the number of users grows larger, i.e., around $1000$, as shown in Fig. \ref{ErrorCorrection}. 


\begin{figure}[t]
	\centering
	\subfigure[Estimation accuracy versus the number of RSS samples with perfect geographical map input.]{\includegraphics[height=4.7cm]{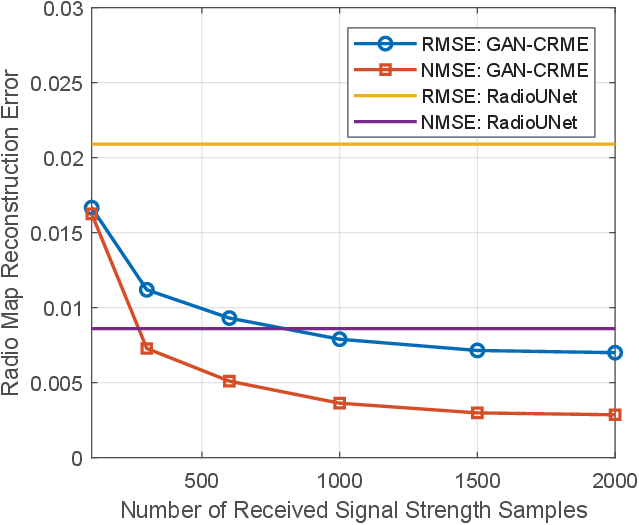}}
	\subfigure[Estimation accuracy versus the number of RSS samples with a random number of missing buildings in the geographical map input.]{\includegraphics[height=4.7cm]{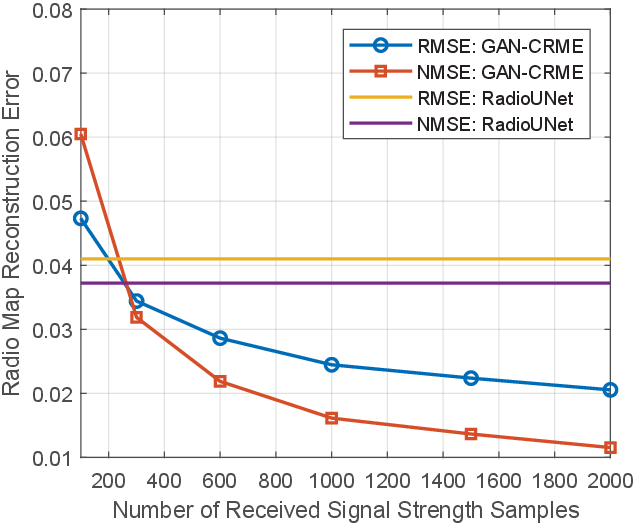}}
	\caption{Effect of number of users on the estimation accuracy.}\label{NumUserEffect}\vspace{-5mm}
\end{figure}


\subsection{Accuracy versus Number of Users}
In this subsection, we show that the accuracy of the proposed GAN-CRME is usually higher than an acceptable threshold and increases proportionally to the number of RSS samples. A comparison of accuracy between the proposed GAN-CRME and the RadioUNet benchmark is given in Fig. \ref{NumUserEffect}, where \emph{normalized-mean-squared-error} (NMSE) and \emph{root-mean-squared-error} (RMSE) are adopted as performance metrics. We can see that GAN-CRME outperforms the RadioUNet when the number of RSS samples exceeds $300$, which can be easily fulfilled in a $256\!\times\!256$ $m^2$ area. Moreover, in view of the fact that the radio map information is more crucial to crowded areas than sparsely populated areas when revisiting the motivation of RME, the proposed GAN-CRME is expected to achieve satisfactory estimation performance in practice.

\section{Conclusion}
In this work, we present GAN-CRME, a fast and environment-aware RME approach. Fast and accurate RME is achieved without the presence of the transmitters' information by exploiting the interaction between distributed RSS measurements and the geographical map information. We also propose a GAN-based learning algorithm, which endows the DNN estimator with high-accuracy RME and error-correction capabilities. With low comutational complexity, low data-acquisition cost and high estimation accuracy, the proposed GAN-CRME would be a promising solution to high-quality RME in different environments with the assist of the geographical map information. The GAN-CRME proposed in this work is helpful for intelligent radio resource management in 6G networks and also sheds light on relevant research topics such as 3D RME, cognitive radio and digital twin.

	\bibliographystyle{ieeetr}
	\bibliography{Ref}

\begin{thebibliography}{10}

\bibitem{CS1}
Y.~Mao, C.~You, J.~Zhang, K.~Huang, and K.~B. Letaief, ``A survey on mobile
  edge computing: The communication perspective,'' {\em IEEE Commun. Surv.
  Tutor.}, vol.~19, pp.~2322--2358, Fourthquarter 2017.

\bibitem{Dingzhu2}
Y.~Shi, Y.~Zhou, D.~Wen, Y.~Wu, C.~Jiang, and K.~B. Letaief, ``Task-oriented
  communications for {6G}: Vision, principles, and technologies,'' {\em IEEE
  Wireless Commun.}, vol.~30, pp.~78--85, Jun. 2023.

\bibitem{Dingzhu1}
D.~Wen, P.~Liu, G.~Zhu, Y.~Shi, J.~Xu, Y.~C. Eldar, and S.~Cui, ``Task-oriented
  sensing, computation, and communication integration for multi-device edge
  {AI},'' {\em IEEE Trans. Wireless Commun.}, early access, 2023.

\bibitem{Application1}
S.~Zhang and R.~Zhang, ``Radio map-based {3D} path planning for
  cellular-connected {UAV},'' {\em IEEE Trans. Wireless Commun.}, vol.~20,
  pp.~1975--1989, Mar. 2021.

\bibitem{VFL}
Z.~Zhang, G.~Zhu, and S.~Cui, ``Low-latency cooperative spectrum sensing via
  truncated vertical federated learning,'' in {\em Proc. 2022 IEEE Globecom
  Workshops (GC Wkshps)}, pp.~1858--1863, Dec. 2022.

\bibitem{Definition1}
D.~Romero and S.-J. Kim, ``Radio map estimation: A data-driven approach to
  spectrum cartography,'' {\em IEEE Signal Process. Mag.}, vol.~39, pp.~53--72,
  Nov. 2022.

\bibitem{Definition2}
K.~Sato, K.~Suto, K.~Inage, K.~Adachi, and T.~Fujii,
  ``Space-frequency-interpolated radio map,'' {\em IEEE Trans. Veh. Technol.},
  vol.~70, pp.~714--725, Jan. 2021.

\bibitem{kriging2}
W.~Van~Beers and J.~Kleijnen, ``Kriging interpolation in simulation: a
  survey,'' in {\em Proc. 2004 Winter Simul. Conf.}, vol.~1, p.~121, Dec. 2004.

\bibitem{kernel}
B.~Schölkopf and A.~Smola, {\em Learning with Kernels: Support Vector
  Machines, Regularization, Optimization, and Beyond}.
\newblock Cambridge, MA, USA: MIT press, 2002.

\bibitem{tensorcompletion}
S.~Shrestha, X.~Fu, and M.~Hong, ``Deep spectrum cartography: Completing radio
  map tensors using learned neural models,'' {\em IEEE Transactions on Signal
  Processing}, vol.~70, pp.~1170--1184, 2022.

\bibitem{Junting1}
W.~Liu and J.~Chen, ``{UAV}-aided radio map construction exploiting environment
  semantics,'' {\em IEEE Trans. Wireless Commun.}, vol.~22, pp.~6341--6355,
  Sept. 2023.

\bibitem{ray}
K.~Rizk, J.-F. Wagen, and F.~Gardiol, ``Two-dimensional ray-tracing modeling
  for propagation prediction in microcellular environments,'' {\em IEEE Trans.
  Veh. Technol.}, vol.~46, pp.~508--518, May 1997.

\bibitem{Junting2}
H.~Sun and J.~Chen, ``Propagation map reconstruction via interpolation assisted
  matrix completion,'' {\em IEEE Trans. Signal Process.}, vol.~70,
  pp.~6154--6169, Dec. 2022.

\bibitem{autoencoder}
Y.~Teganya and D.~Romero, ``Deep completion autoencoders for radio map
  estimation,'' {\em IEEE Trans. Wireless Commun.}, vol.~21, pp.~1710--1724,
  Mar. 2022.

\bibitem{Unet}
R.~Levie, C.~Yapar, G.~Kutyniok, and G.~Caire, ``{RadioUNet:} fast radio map
  estimation with convolutional neural networks,'' {\em IEEE Trans. Wireless
  Commun.}, vol.~20, pp.~4001--4015, Jun. 2021.

\end{thebibliography}
\end{document}